\documentclass[aps, twocolumn, prx, showpacs, preprintnumbers, amsmath, amssymb, superscriptaddress]{revtex4-2}
\usepackage{amsmath, amsfonts, amssymb, graphics, graphicx, epsfig, color, times, indentfirst, layout, hyperref, subfigure, multirow, bm}
\usepackage{hyperref, soul}
\usepackage{ulem}
\hypersetup{linktocpage,,colorlinks=true}
\usepackage{threeparttable}
\usepackage{dcolumn}
\usepackage{bm}
\usepackage{color}

\begin{document}

\title{Disorder effects on triple-point fermions}

\author{Hsiu-Chuan Hsu}\email{hcjhsu@nccu.edu.tw}
\affiliation{Graduate Institute of Applied Physics, National Chengchi University, Taipei 11605, Taiwan}
\affiliation{Department of Computer Science, National Chengchi University, Taipei 11605, Taiwan}

\author{Ion Cosma Fulga}
\affiliation{Institute for Theoretical Solid State Physics, IFW Dresden and W{\"u}rzburg-Dresden Cluster of Excellence ct.qmat, Helmholtzstr. 20, 01069 Dresden, Germany}

\author{Jhih-Shih You}\email{jhihshihyou@ntnu.edu.tw}
\affiliation{Department of Physics, National Taiwan Normal University,  Taipei 11677, Taiwan}

\date{\today}
\begin{abstract}
The stability of three-dimensional relativistic semimetals to disorder has recently attracted great attention, but the effect of disorder remains elusive for multifold fermions, that are not present in the framework of quantum field theory. 
In this paper, we investigate one type of multifold fermions, so-called triple-point fermions~(TPFs), which have pseudospin-1 degrees of freedom and topological charges $\pm2$. 
Specifically, we consider the effect of disorder on a minimal, three-band tight-binding model, which realizes the minimal number of two TPFs. 
The numerically-obtained, disorder-averaged density of states suggests that, within a finite energy window, the TPFs are robust up to a critical strength of disorder. 
In the strong disorder regime, the inter-TPF scattering is the main mechanism for destroying a single TPF. 
Moreover, we study the effects of disorder on the distribution of Fermi arcs and surface Berry curvature. 
We demonstrate that the Fermi arc retains its sharpness at weak disorder, but gradually dissolves into the metallic bulk for stronger disorder. 
In clean limit, the surface Berry curvature exhibits a bipolar configuration in the surface Brillouin zone. 
With increasing disorder, the positive and negative surface Berry curvature start to merge at the nearby momenta where the Fermi arcs penetrate into bulk.
\end{abstract}

\maketitle

\section{Introduction}
Fundamental particles in high-energy physics can emerge in condensed-matter systems. 
For example, Majorana fermions could be realized in topological superconductors~\cite{Qi2011c}, Dirac fermions in graphene~\cite{CastroNeto2009a}, and Weyl fermions in topological semimetals~\cite{Armitage2018}. 
These aforementioned quasiparticles are described by two-component spinors. 
Nonetheless, it has been shown that, with particular crystal symmetries, lattice systems could host new types of quasiparticles with higher pseudospin~\cite{Bradlyn2016, Hasan2021}, which have no high-energy counterparts. 
One of these exotic quasiparticles is realized as a low energy excitation at a point where three energy bands become degenerate. 
The resulting three-fold fermion has pseudospin-1 degrees of freedom, and has been dubbed the triple point fermion~(TPF)~\cite{Bradlyn2016}. 
TPFs have topological charge of $\pm2$, the same as that of double Weyl nodes. 
However, a TPF has three energy bands, two of which disperse linearly along any momentum direction, and one of which has a vanishing velocity at the band crossing point, i.e., it is a (locally) flat band.
In crystals, a stable topological charge is separated from its partners with opposite charge in the Brillouin zone, since the total charge must vanish, as dictated by the Nielsen-Ninomiya theorem \cite{Nielsen1981a, Nielsen1981b}. 
When two nodes of opposite topological charge are mixed, for instance by inter-node scattering, their topologically nontrivial properties are no longer guaranteed, and can even disappear altogether. 

The last few years have witnessed an enormous interest to understand the stability of the Weyl nodes against disorder~\cite{Kobayashi2014,Sbierski2014, Chen2015, Goswami2015, Bera2016, Pixley2016, Sbierski2017, Klier2019, Wilson2020, Pixley2021,Pires2021,Pires2022}. 
The disorder-averaged density of states~(DOS) has been used as an order parameter to characterize the stability of Weyl nodes. 
Within self-consistent Born approximation, it has been shown that there exists a critical disorder strength that separates the Weyl semimetal and diffusive metal regimes~\cite{Sbierski2017, Klier2019}.
Meanwhile, numerical simulations performed using the kernel polynomial method revealed that the Weyl node is unstable, and the DOS becomes finite in the presence of disorder due to the rare region effects~\cite{Nandkishore2014, Pixley2016, Wilson2020}. 
The stability of quadratic Weyl semimetals has also been investigated. 
It has been shown that linear Weyl semimetals are stable against random disorder, but the quadratic ones are not~\cite{Bera2016}. 
Similarly, Sbierski~\textit{et~al.}~\cite{Sbierski2017} found that double Weyl nodes are unstable against correlated disorder.
While a TPF shares the same topological charge with the double Weyl point, $\pm2$, the disorder effects in TPFs are still overlooked.
It is an open question whether richer and more interesting physics will emerge in the TPFs when disorder and the additional flat band come into play.

A direct consequence of the topological charges is the existence of Fermi arcs connecting the surface projections of oppositely-charged nodes. 
These Fermi arcs are manifestations of the bulk-boundary correspondence of Weyl semimetals and can be measured by ARPES~\cite{Armitage2018} or in transport experiments~\cite{Hu2019}. 
Moreover, the Fermi arcs carry surface Berry curvature~(BC) and Berry curvature dipole, leading to linear or nonlinear anomalous Hall conductivity~\cite{Chen2013, Wawrzik2021}. 
So far, there are only few works devoted to studying the robustness of the Fermi arcs of Weyl semimetals~\cite{Takane2016, Slager2017, Wilson2018}. 
Even less is known about the disorder effects on the Fermi arcs of TPFs.





Due to the additional flat band, the DOS, a commonly
used diagnostic that indicates the stability of Dirac/Weyl nodes against disorder, cannot solely reveal the disorder effects on TPFs. Therefore,  in addition to the DOS, we study the fate of Fermi arcs as well as the associated BC driven phenomena.
This work is organized as follows. In Sec.~\ref{sec:model}, we introduce the model Hamiltonian of the triple point fermion. 
The numerical results are presented and discussed in the following sections: Section \ref{sec:dos} shows the results of density of states and Sec.~\ref{sec:arcs} discusses the stability of Fermi arcs and the interplay between disorder and the surface Berry curvature. 
Finally, we conclude in Sec.~\ref{sec:concl}.

\section{model Hamiltonian}\label{sec:model}

In this study, we use the minimal lattice model that contains only a pair of triple point fermions~\cite{Fulga2017}. 
This Hamiltonian, which is constructed as a combination between a double Weyl semimetal and a flat band at zero energy, reads
\begin{eqnarray}
H_t(\vec{k})=\begin{pmatrix}
H_q&& \lambda_+^{\dagger}\\
&&\lambda_-^{\dagger}\\
\lambda_+& \lambda_-&0
\end{pmatrix},
\label{eq:htpf}
\end{eqnarray}
where $H_q$ is the three-dimensional (3D) momentum-space Hamiltonian for the double Weyl semimetal,
\begin{eqnarray}
H_{q}(\vec{k})=\left[2-\cos(k_x)-\cos(k_y)-2\cos(k_z)
\right]\sigma_z +\nonumber\\
2\sin(k_x)\sin(k_y)\sigma_y+
2\left[\cos(k_x)-\cos(k_y)
\right]\sigma_x,
\label{eq:hq}
\end{eqnarray}
$\sigma_j$ are Pauli matrices, $\vec{k}=(k_x, k_y, k_z)$ is the momentum vector, and $\lambda_{\pm}=\lambda e^{i(\phi\pm\pi/4)}(\sin (k_x)\mp i\sin(k_y))$ are the coupling terms between the double Weyl semimetal and the flat band. 
Hereafter, we set $\hbar=1$ and the hopping energy and lattice constant to $1$ for simplicity.

\begin{figure}[tb]
  	\includegraphics[width=0.4\textwidth]{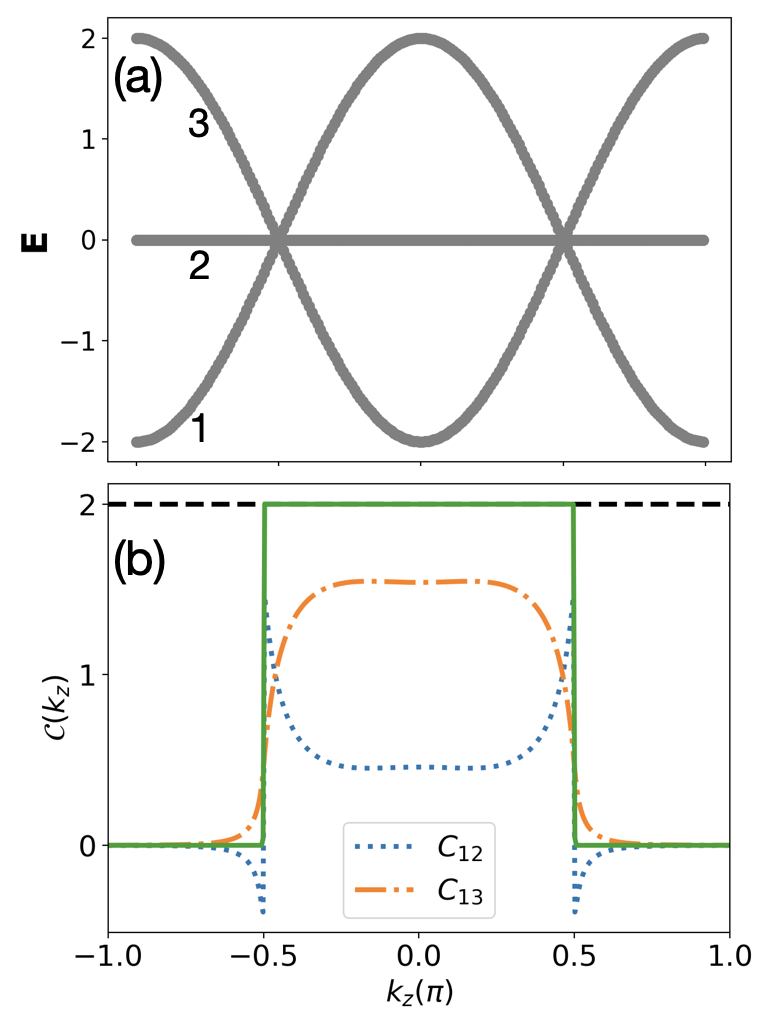}
  	\caption{ 
  	(a) The energy dispersion for $\lambda=0.5$ along $k_z$ at $(k_x, k_y)=(0,0)$. 
  	(b) The Chern number (green solid line) along $k_z$. 
  	The blue dotted (orange dashed dotted) line labeled by '12(3)' denotes the contribution to the Chern number obtained from the matrix element between the valence and flat (conduction) band. The dashed black line is a guide to the eye that shows the quantization of the total Chern number.
  	}
  	\label{fig:chern}
  \end{figure}

For the double Weyl semimetal $H_q$, the two bands near the Weyl nodes $(0,0,\pm\pi/2)$ disperse linearly along $k_z$, but quadratically along $k_x$ and $k_y$. 
By introducing a flat band that couples to the double Weyl fermions, we obtain a triple point fermion~\cite{Fulga2017}, such that two of the bands disperse linearly in all momentum directions. 
The dispersion for $\lambda=0.5$ along $k_z$ at $(k_x, k_y)=(0,0)$ is shown in Fig.~\ref{fig:chern}(a). 
By Fourier transforming to real-space, we obtain a tight-binding Hamiltonian which allows us to explore the effects of disorder. 

The band crossings of $H_t$ and $H_q$ have the same topological charge \cite{Nandy2019}, and the TPFs are protected by the combination of an anticommuting mirror symmetry about the $k_x=k_y$ plane and a $C_4$ rotation symmetry along the $k_x=k_y=0$ axis \cite{Fulga2017}. 
This is the minimal model for TPFs with a symmorphic group \cite{Fulga2017}.
Moreover, $H_t$ and $H_q$ are time-reversal symmetry broken and can be viewed as a stack of quantum anomalous Hall insulators along the $k_z$ direction. 
The Chern number along $k_z$ for the $n^{\rm th}$ band is given by
\begin{eqnarray}
\mathcal{C}^n(k_z)&=&\frac{1}{\pi} \int dk_xdk_y \Omega^{n}_z(\vec{k}),
\end{eqnarray}
where $\Omega^{n}_z(\vec{k})$ is the z-component of the Berry curvature for the $n^{\rm th}$ band
\begin{eqnarray}
\Omega_z^{n}(\vec{k})&=&
-{\rm Im} \sum_{n\neq m}\frac{\langle n|v_x|m\rangle\langle m|v_y|n\rangle}
{(E_n-E_m)^2},
\end{eqnarray}
where $E_n$ and $|n\rangle$ are eigenenergies and Bloch eigenstates of the Hamiltonian $H_t(\vec{k})$, respectively, and $v_i=\partial H_t(\vec{k})/\partial k_i$ is the velocity operator.
For $H_t$, the flat band does not carry a Chern number, while the lowest and highest energy bands carry opposite Chern numbers. 
For the lowest energy band~(band 1), there are two contributions to the Chern number. 
One is from the velocity matrix elements between the band 1 and 3, denoted by $C_{13}$ and the other is between the band 1 and 2, denoted by $C_{12}$. 
The Chern number, Eq.~\eqref{eq:htpf}, of the lowest energy band at each $k_z$ plane is shown by the green solid line in Fig.~\ref{fig:chern}(b). 
It is well-quantized between the two nodes. 
Both matrix elements, 12 and 13, are nonzero for $\lambda=0.5$. 
In contrast, when $\lambda=0$, the topological charge is carried only by the double Weyl node, which is now decoupled from the flat band, and only the $C_{13}$ component is nonzero. 
Therefore, as the coupling strength $\lambda$ increases, there is a Berry curvature transfer between the bands, as shown in the Appendix \ref{app:bctransfer}. 

\section{Density of states}\label{sec:dos}

A 3D relativistic semimetal has the low-energy dispersion, $E=\pm v_f k$, where $v_f$ is the slope of the band and $k=\sqrt{k_x^2+k_y^2+k_z^2}.$
In the clean limit, it is straightforward to show that the DOS is
$\rho_{\rm clean}(E)=\frac{\pi^2}{2}k^2\left(
\frac{dk}{dE}
\right)^{-1} \nonumber = \frac{\pi^2}{2 v_f^3}E^2$.
In the presence of disorder, the disorder-averaged DOS $\rho(E)$ at zero energy can serve as an order-parameter for the disorder-driven semimetal-metal transition.
For TPFs, band 1 and 3 disperse linearly and thus the DOS of both bands still scales quadratically with energy~\cite{Pal2022}. However, due to the existence of the flat band, the DOS has an additional zero-energy peak. 
This is in stark contrast to the DOS of Weyl and Dirac fermions, and distinct phenomena arise from the TPFs with disorder.

To calculate the disorder-averaged DOS, we employ the kernel polynomial method \cite{Weisse2006} in the pybinding \cite{Moldovan2020} package. 
Two types of disorder are considered. 
The first type is the white noise, which takes into account the scattering between the two TPFs.
We add a random onsite potential, $V(\vec{r}) I_{3\times 3}$, with $I_{3\times 3}$ the three by three identity matrix, and $V(\vec{r}) \in [\frac{-W}{2},\frac{W}{2} ]$ drawn independently for each lattice site (indexed by real-space position vectors $\vec{r}$) from a uniform distribution, where $W$ is the strength of disorder. 

The second type is the correlated disorder,
\begin{eqnarray}
H_c&=&\sum_{i}\sum_j\frac{W_j}{(\sqrt{2\pi}\xi)^3}e^{-\frac{(r_i-R_j)^2}{2\xi^2}} I_{3\times 3}|i\rangle\langle i|,
\end{eqnarray}
where $\xi$ is the correlation length, $W_j$ is the random number given by a uniform distribution in the range $\in [\frac{-W}{2},\frac{W}{2} ]$, $r_i$ is the position vector for a lattice site and $R_j$ is the position vector of an impurity. It is assumed that the number of impurity is the same as that of the lattice sites.
When $\xi\rightarrow 0$, the correlated disorder reduces to the white noise potential.
As the correlation length $\xi$ becomes larger, the scattering between two TPFs becomes weaker~\cite{Sbierski2017}. 

\begin{figure}[tb]
	\includegraphics[width=0.45\textwidth]{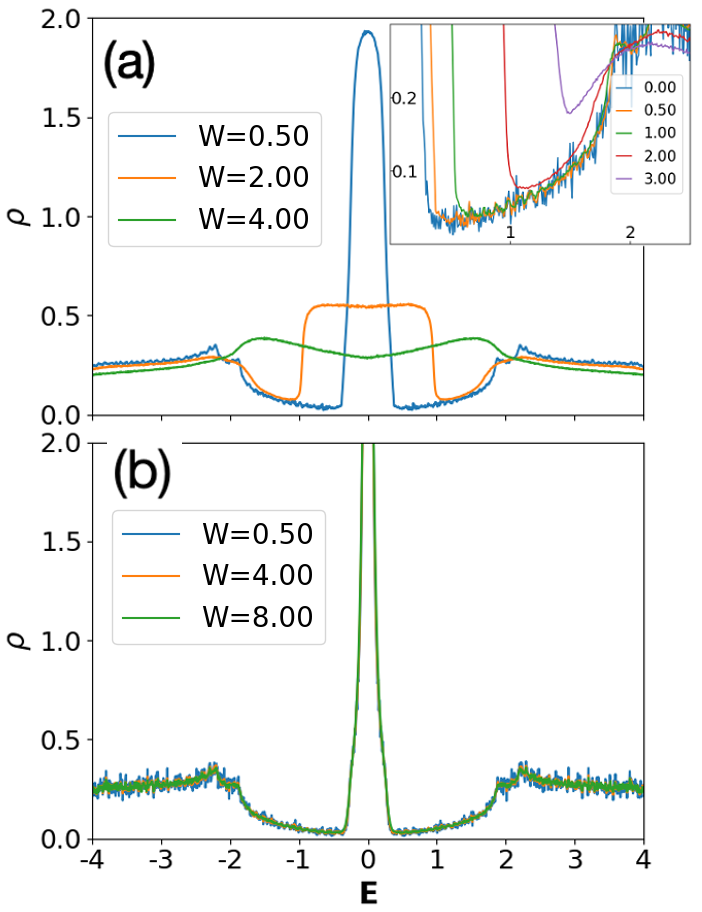}
	\caption{
	The DOS as a function of energy for $\lambda=0.5$, for a cubic system with a linear size of $L=50$ lattice sites. A Gaussian broadening of $0.001$ was used when implementing the kernel polynomial method, and the DOS was averaged over 20 disorder realizations. 
	(a) DOS for the white noise disorder. 
	The inset is the zoomed-in plot showing the quadratic behavior of the DOS. 
	(b) DOS for the correlated disorder with $\xi=5$.
	}
	\label{fig:dos}
\end{figure}

Figure~\ref{fig:dos}(a) shows the results for white noise disorder, computed for a cubic system consisting of $L^3$ sites, with $L=50$ and periodic boundary conditions imposed along all three directions. 
The zero-energy peak decreases and broadens as the disorder strength increases. 
As can be seen, there exists an energy range for which the DOS remains quadratic up to $W\leq 3$. 
In contrast, the double Weyl nodes have a linear DOS at finite energy in the presence of disorder~\cite{Goswami2015, Bera2016, Sbierski2017}.
Figure \ref{fig:dos}(b) shows the results for correlated disorder, using the same geometry.
In contrast to white noise disorder, the DOS is almost invariant to the disorder strength. 
This result suggests that the inter-TPF scattering is the main mechanism for destroying the single TPF physics. 

We note that the TPF and multifold fermions have been predicted in a family of transition metal silicides, in which the multifold fermions are separated in energy~\cite{Tang2017}. 
Effectively, there is only one multifold fermion involved in scattering events. 
Thus, our results suggest that such a multifold fermion in this material family should be particularly robust against disorder. 

\section{Fermi arcs}\label{sec:arcs}

\begin{figure}[tb]
	\includegraphics[width=0.5\textwidth]{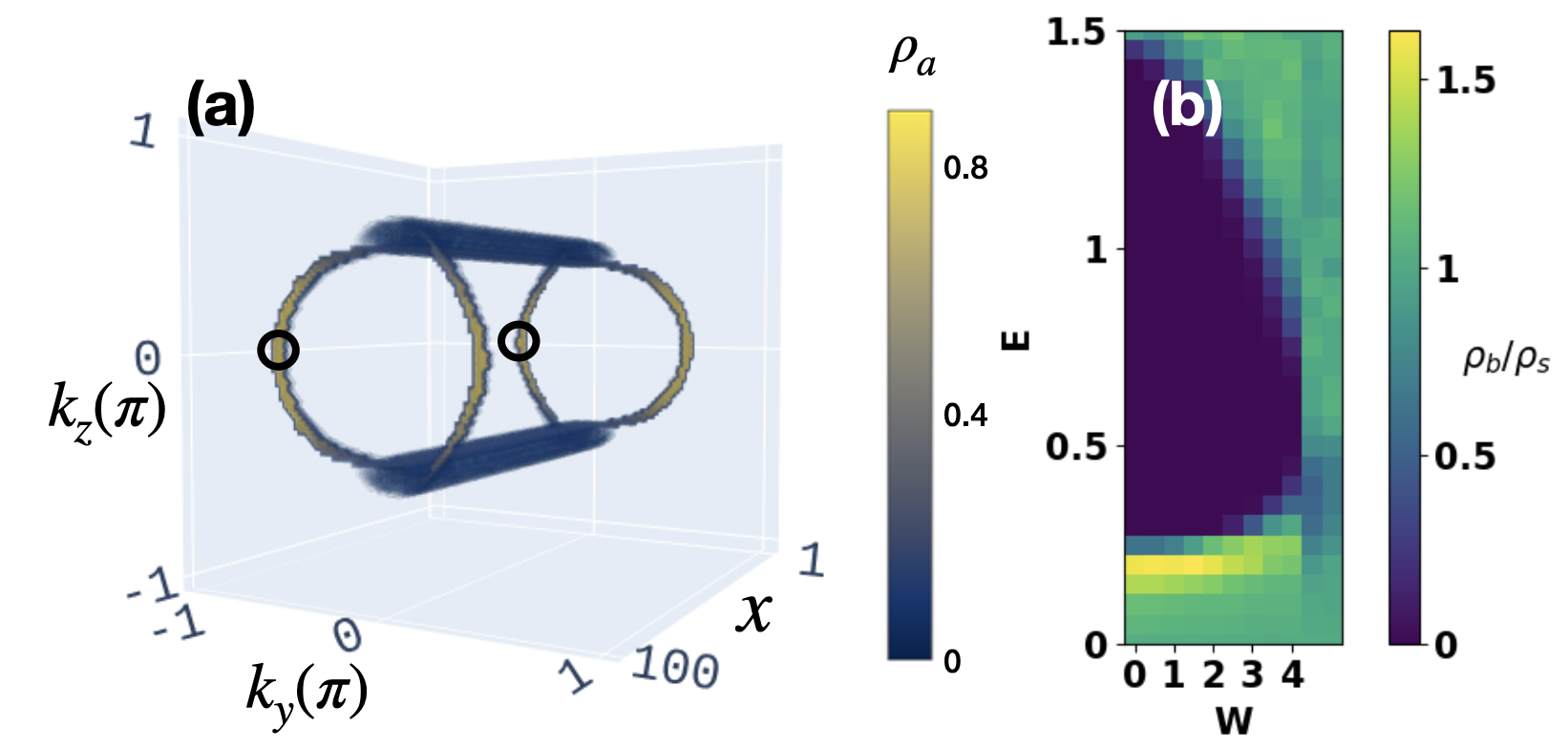}
	\caption{ 
	(a) Distribution of states in the energy range $0.3 \leq E < 0.4$ for $\lambda=0.5$, in the absence of disorder. {The black circles denote two of the furthest momenta away from the TPFs.}
	(b) The ratio of the bulk and surface density near $k'$ in a slab geometry. 
	The number of lattice sites of the slab is $100$ along the x-direction, and we have averaged over 10 disorder realizations.
	}
	\label{fig:phase}
\end{figure}

The existence of Fermi arcs is a consequence of the topological charges associated to the band crossing points. 
Several measurable quantities are related to the Fermi arcs and thus the stability of the Fermi arcs against disorder is an essential question. 
To study the topological surface states associated with the TPFs, we impose open boundary conditions along the x-direction of the model Hamiltonian, while keeping the system infinite along the other two directions, such that $k_{y,z}$ remain good quantum numbers. 
Since the central band is close to $E=0$ and hinders the observation of Fermi arcs, here we focus on the Fermi arc state in a finite energy range away from zero.
The density of the Fermi arc states $(\rho_a)$ in the clean limit is shown in Fig.~\ref{fig:phase}(a). Here $\rho_a$ is defined as 
\begin{equation}\label{eq:rho_a}
\rho_a(x, k_y, k_z) = \sum_{E_i \in \left[0.3, 0.3+\Delta E \right)}\sum_{o=1}^3| \langle \psi_i(k_y,k_z) | x,o\rangle|^2,
\end{equation}
where $\psi_i(k_y, k_z)$ is the $i^{\rm th}$ eigenvector with energy $E_i \in \left[0.3,0.3+\Delta E\right)$ at $(k_y, k_z)$, $x$ denotes the coordinate of lattice sites along the finite direction of the slab, and $o$ labels the pseudospin degrees of freedom.
{ Here $\Delta E$ was set to be $0.1$ to ensure the continuity of the Fermi arcs in the projected Brillouin zone in Fig.~\ref{fig:phase}(a). The value could be smaller if we use a finer momentum grid. 
}
On each surface, there exists a pair of Fermi arcs connecting to the bulk states of TPFs at $(k_y=0, k_z=\pm \pi/2)$. 
The number of Fermi arcs is consistent with the fact that TPFs are monopoles of the Berry curvature with charge $\pm2$. 

\begin{figure}[tb]
	\includegraphics[width=0.4\textwidth]{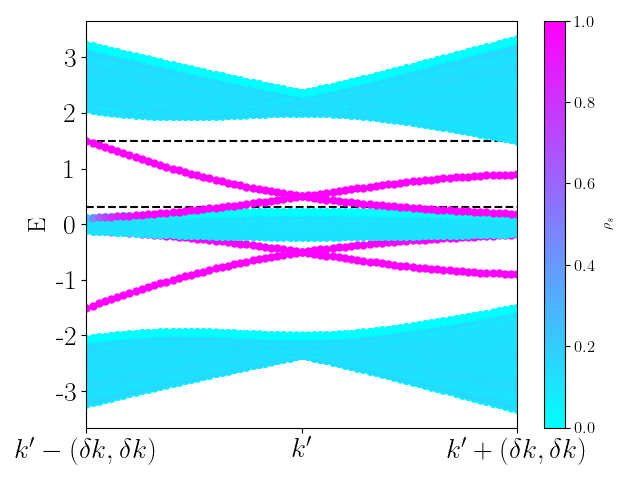}
	\caption{ 
		Energy spectrum of the slab. The horizontal black dashed lines denotes $E=0.3$ and $E=1.5$. There are 100 lattice site along the x-direction. The color on the curved represents surface density. $k'=(k_y=-\pi/2,k_z=0)$ and $\delta k=10\frac{2\pi}{N}, \ N=100$.}
	\label{fig:slabdisp}
\end{figure}

To investigate the robustness of the Fermi arcs, we employ the stacked-layer construction~\cite{Slager2017}, in which the disorder potential is applied only in the x-direction~(See Appendix~\ref{app:slconstruction}). 
This allows us to study the evolution of the Fermi arcs with disorder and in turn to compute the local density near the momentum $k'=(k_y=-\pi/2,k_z=0),$ which is one of the furthest momenta away from the TPFs. 
We choose this k-point because the surface Fermi arcs at this point can have the least overlap with the bulk contributions of TPFs. 
{The dispersion for the slab near $k'$ is shown in Fig.~\ref{fig:slabdisp}. It is plotted along the diagonal that runs from $k'-(\delta k,\delta k)$ to $k'+(\delta k,\delta k)$, where $\delta k=\pi/5$. The flat band shows a finite band width. Thus, in this range of momentum, the Fermi arcs can only be isolated from bulk for $0.3 < E<1.5$. The highest energy states of the Fermi arcs (megenta curves) eventually merge with bulk states (cyan curves) when momentum is away from $k'$.}

\begin{figure*}[tb]
	\includegraphics[width=0.9\textwidth]{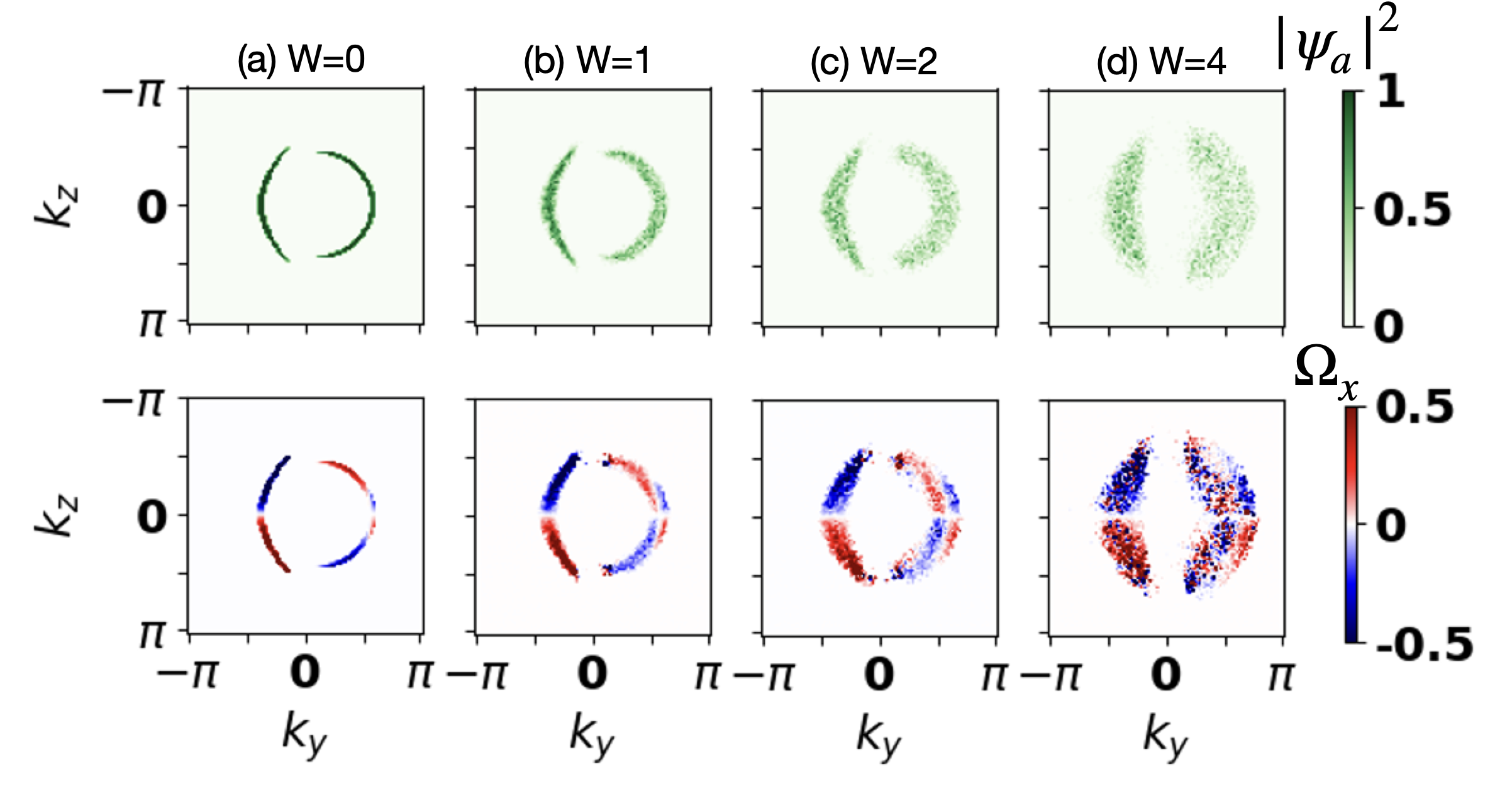}
	\caption{
	Fermi arcs (top panel) and the Berry curvature carried by the Fermi arcs (lower panel) on the top surface for disorder (a) $W=0$, (b) $W=1$, (c) $W=2$ and (d) $W=4$. For (b-d), we have avraged over 20 disorder realizations.
	At stronger disorder the distribution of Fermi arcs is rather sparse. 
	According to Fig.~\ref{fig:phase}(b), the Fermi arcs have penetrated into bulk at $W=4$.
	}
	\label{fig:surfbc}
\end{figure*}

When the disorder strength increases, the Fermi arcs at $k'$ gradually penetrate into bulk. 
We calculate the ratio of the bulk~($\rho_b$) and surface~($\rho_s$) density to quantify this disorder-induced penetration as a function of energy and disorder strength. The definition of $\rho_{b/s}$ is
 \begin{eqnarray}
 \rho_{b/s}=\sum_{k_y,k_z}\sum_i
 {\rho^i_{b/s}(k_y,k_z)},
 \label{eq:rhobs}
 \end{eqnarray}
where the first sum is performed over the neighboring k-points near $k'$, within the range of $\left[ k_y'-\delta k,k_y'+\delta k\right]$ and $\left[ k_z'-\delta k,k_z'+\delta k\right]$. Here we choose $\delta k=10\frac{2\pi}{N},$ where $N=100$ is the number of k-points along each direction.  {Furthermore, $i$ denotes the states within the energy range $E\leq E_i < E+\Delta E$, where $\Delta E$ was set to be $0.1$ as for the Fermi arcs shown in Fig.~\ref{fig:phase} (a). $\rho_{b/s}^i(k_y,k_z)$ is defined as }
 \begin{eqnarray}
 \rho_{b}^i(k_y,k_z)&=&\sum_{x=d}^{L-d}\sum_{o=1}^3|\langle \psi_{i}(k_y,k_z)|x,o\rangle|^2, \\
 \rho_{s}^i(k_y,k_z)&=&1-\rho^i_b(k_y,k_z),
 \end{eqnarray}
where $L=100$ is the thickness of the slab and $d=L/20$ is the number of layers that are considered as top/bottom surfaces. 
The diagram showing this ratio is given in Fig.~\ref{fig:phase} (b). 
In the clean limit, i.e. $W=0$, the bulk density vanishes for $0.3<E<1.5$, the energy range where the Fermi arcs are isolated from the bulk states. 
Outside this range, the bulk states coexist with Fermi arcs near $k'$. 
As $W$ becomes larger, the Fermi arcs gradually penetrate into bulk and $\rho_b$ increases. 
The Fermi arcs can carry a large surface Berry curvature $\Omega_{x}(\vec{k})$~\cite{Wawrzik2021}. 
As is well-known, in the presence of an external electric field $\vec{E}$, the Berry curvature in momentum space adds the so-called anomalous velocity term in the equations of motion that describe the propagation of Bloch wave-packets, $-\frac{e}{\hbar} \vec{E} \times \vec{\Omega}(\vec{k})$~\cite{Xiao2010, Fuchs2010, Xu2014}.
Since the anomalous velocity is perpendicular to the electric field, it can lead to a variety of Hall-like effects. 
We calculate the surface BC for the Fermi arcs on the top surface of the slab and investigate the effect of disorder. The Fermi arcs are the states with the largest weight on the surface, denoted by $\psi_a$. 
Figure~\ref{fig:surfbc} shows the Fermi arcs and the surface BC at $E=0.3$ for $W=0, 1,2,4$. 
In the clean limit, we find that BC of each arc is odd in $k_z$, $\Omega_{x}(k_y, k_z)=-\Omega_{x}(k_y,-k_z)$.
Such a bipolar configuration of BC is due to the mirror symmetry in the z-direction, i.e. $H(-k_z)=H(k_z)$, according to Eqs.~\eqref{eq:htpf} and \eqref{eq:hq}. 
The BCs on the top~(t) and bottom~(b) surfaces are related by the inversion symmetry of the slab, $\Omega_{x,t}(k_y, k_z)=\Omega_{x,b}(-k_y,-k_z)$. 
Since the BC is proportional to the matrix element of $\frac{\partial{H}}{\partial k_z},$ the Berry curvature is odd in $k_z$. 
Therefore, the surface Hall conductivity, which is proportional to the sum of the Berry curvature in the surface BZ, is exactly zero.


 
Figure~\ref{fig:surfbc} (b) shows that, for relatively weak disorder ($W=1$), the Fermi arcs retain their sharpness. 
However, negative (positive) surface Berry curvature arises near the surface projection of the positively charged (negatively charged) bulk node. 
The mixture of positive and negative surface Berry curvature is more obvious as disorder strength increases, as shown in Fig.~\ref{fig:surfbc}(c) and (d). 
We note that the $k_z$ reflection symmetry is broken and thus the Berry curvature is no longer an odd function in $k_z$ when the potential disorder has distribution in the $k_y$ and $k_z$ direction in the momentum space. 

Finally, we remark that even when the Hall conductivity vanishes in experiments, the intrinsic geometry of the quantum electron wavefunctions can still be probed via the nonlinear Hall effect, a direct measure of the Berry curvature dipole. 
In our case the surface BC dipole defined as 
\begin{eqnarray}
D_i = - \int d^2 \vec{k} \Omega_{x}(\vec{k}) [\partial_{k_i} f_0 (\vec{k}) ],
\end{eqnarray}
where $f_0 (\vec{k})$ is the Fermi-Dirac distribution.
The above equation describes the first order moment of the Berry curvature over occupied states near the Fermi energy.
When the boundary is open along the x-direction, the mirror symmetry in the z-direction forces the surface BC dipole to be $D_{y}= 0$ and $D_{z}\neq0$ ~(see Appendix~\ref{app:bcd}). 
This results in a transverse charge current along the y-direction in second-order response to an external electric field along the z-direction. 
By using the stacked-layer construction, we also numerically find the surface BC dipole could be enhanced with disorder. 
The interplay between disorder and the quantum geometric aspects of the Fermi-arcs' electronic structure is experimentally relevant in all BC driven phenomena, such as the nonlinear anomalous Nernst effect~\cite{Yu2019} and the surface contributions to the orbital magnetic moment~\cite{Ganesh1999}. 
These are certainly interesting subjects for future investigation.

\section{Conclusion}\label{sec:concl}

We have studied the DOS and Fermi arcs of a disordered, three-band tight-binding model that hosts TPFs. Using the lattice model, rather than the low energy model, we are able to investigate the effect of internode scattering. 
We found that the inter-TPF scattering is the main mechanism for destroying the single TPF physics. 
Nonetheless, there exists a finite energy window in which the Fermi arcs and the surface Berry curvature remain sharp in the weak disorder limit. 
These results suggests that the TPFs in the family of transition metal silicides are robust against disorder. 
Furthermore, we show that even when the Hall conductivity carried by the Fermi arc states is zero due to mirror symmetry, a finite surface Berry curvature dipole can still exist, leading to the nonlinear Hall effect.  
The disorder effects on the nonlinear responses such as the nonlinear anomalous Nernst effect and photocurrent would be interesting directions for future research.

 Lastly, we would like to note that the realization of TPFs has also been proposed in optical lattices~\cite{Fulga2018}. In ultracold atomic systems,
disorder can be  implemented by using laser speckle patterns~\cite{Bloch2008}.
In addition, the spectrum can be probed using the methods of time-of-flight
imaging and Bragg spectroscopy~\cite{Bloch2008} and the surface Berry curvature can be mapped by extending the techniques applied for two-dimensional lattices~\cite{Flaschner2016}. Thus, the disorder effects on TPFs  discussed in the paper could be investigated using ultracold atoms in optical lattices.

\section*{acknowledgments}
H.C.H. is supported by the Ministry of Science and Technology (MOST) in Taiwan under the grant MOST 110-2112-M-004-001. J.-S. Y. is supported by the Ministry of Science and Technology, Taiwan (Grant No. MOST 110-2112-M-003-008-MY3) and National Center for Theoretical Sciences in Taiwan.
I.C.F. acknowledges support from the Deutsche Forschungsgemeinschaft (DFG, German Research Foundation) under Germany’s Excellence Strategy
through the W{\"u}rzburg-Dresden Cluster of Excellence on
Complexity and Topology in Quantum Matter – ct.qmat
(EXC 2147, project-id 390858490).

\appendix

\section{Berry curvature transfer}\label{app:bctransfer}

\begin{figure}[tb]
	\includegraphics[scale=0.35]{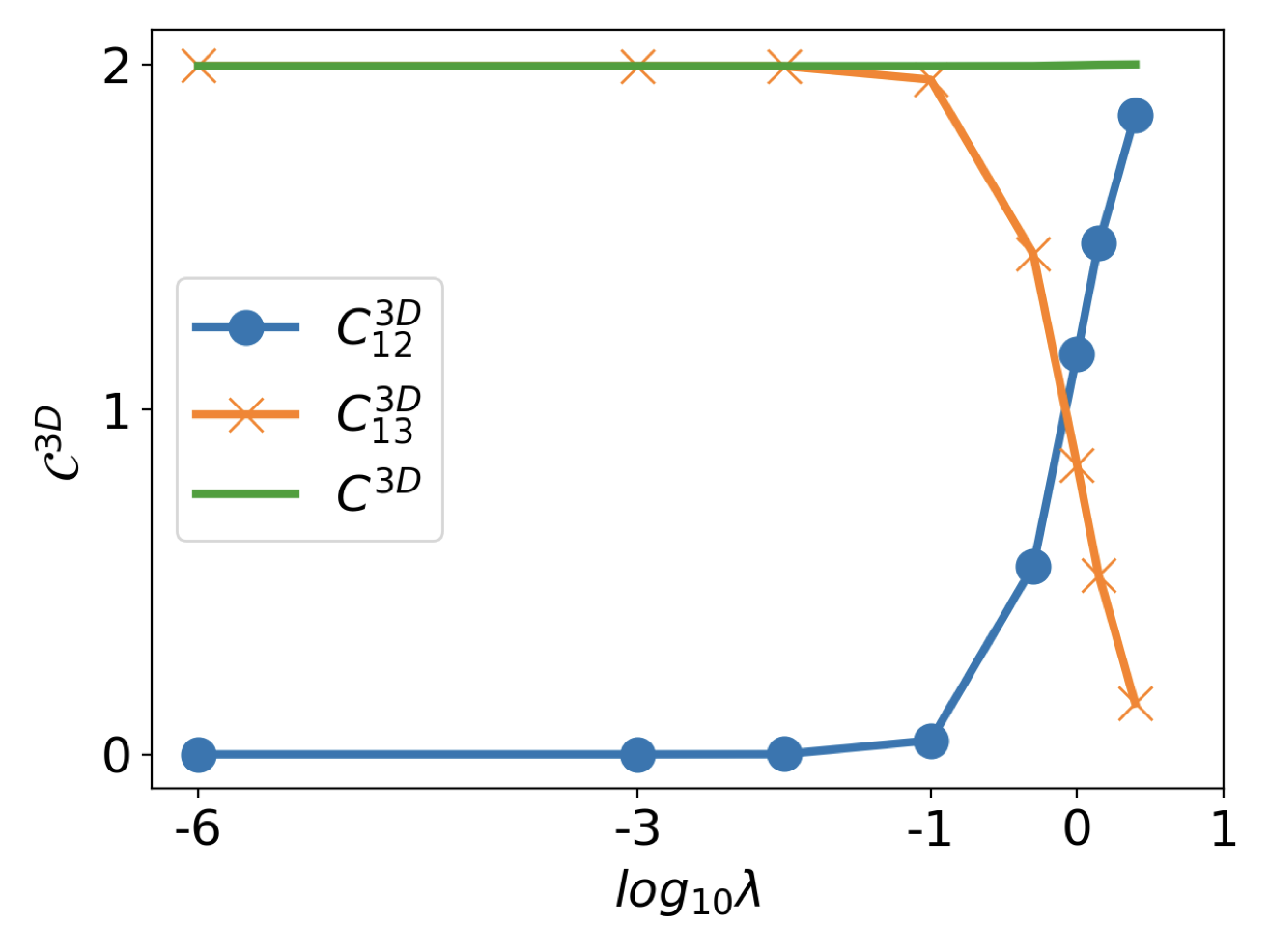}
	\caption{
	The accumulated Chern number for the lowest energy band given by Eq.~\eqref{eq:3dchern}. 
	The Berry curvature transfers from the element between band 1 and 3 to band 1 and 2 as the coupling strength increases. 
	}
	\label{fig:bctransfer}
\end{figure}

The accumulated Chern number for the lowest energy band ($C^{\rm 3D}$), contributed from each $k_z$ layer, is given by~\cite{Burkov2011a}
\begin{eqnarray}
C^{\rm 3D}=\frac{1}{\pi}\int_{k_1}^{k_2}d k_z C(k_z),
\label{eq:3dchern}
\end{eqnarray}
where $k_{1,2}$ are the two TPF nodes. 
$C^{\rm 3D}$ as a function of $\lambda$ is shown in Fig.~\ref{fig:bctransfer}. 

\begin{figure*}[tb]
	\includegraphics[width=0.9\textwidth]{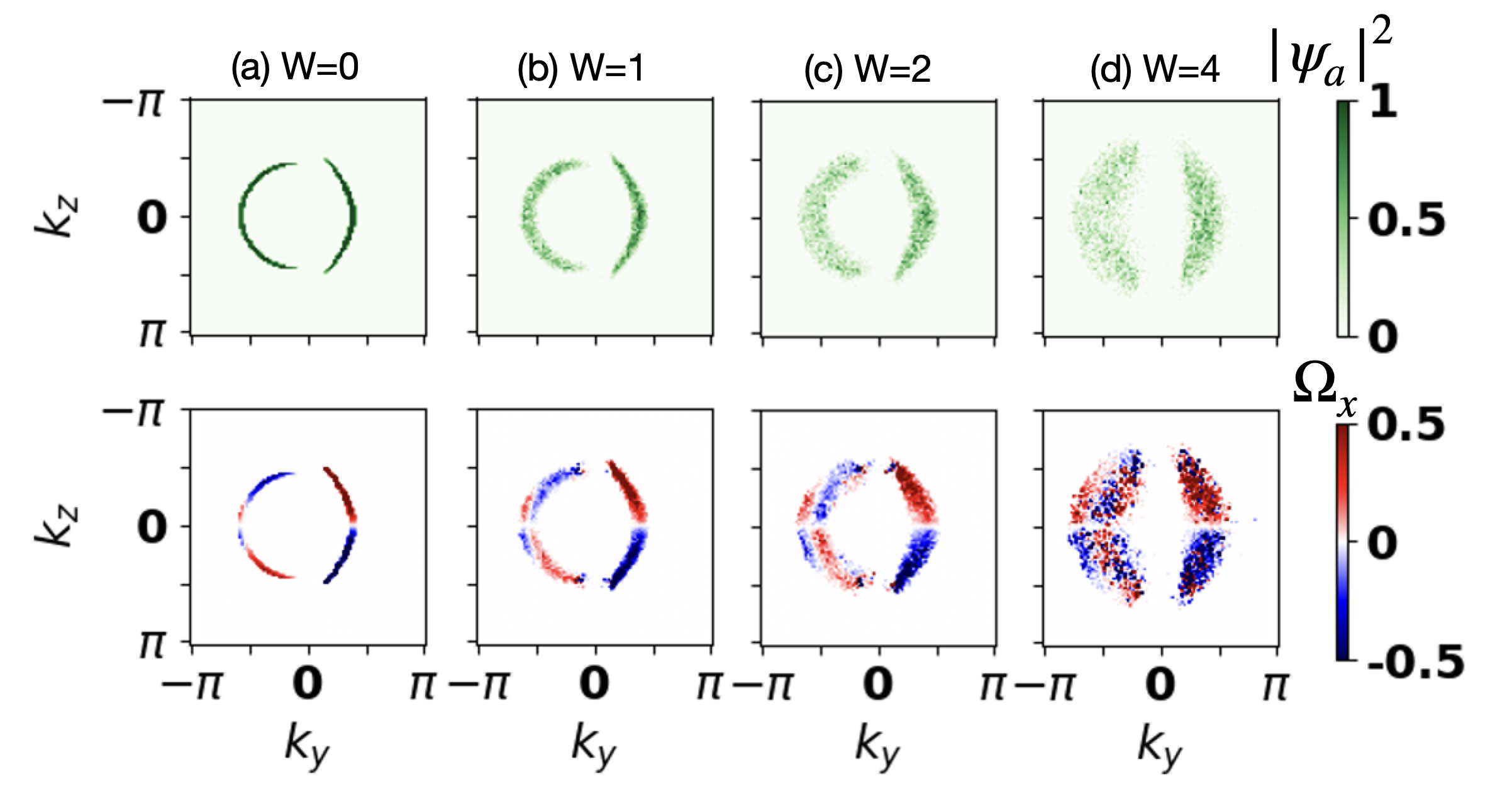}
	\caption{ 
	The same plots as in Fig.~\ref{fig:surfbc}, for the bottom surface of the slab.
	}
	\label{fig:surfbcb}
\end{figure*}

\section{Stacked-layer construction}\label{app:slconstruction}

Here we review the recently-introduced stacked-layer construction~\cite{Slager2017} which is used to study the robustness of the Fermi arcs against disorder.
We consider a slab geometry, infinite along the y and z directions, and impose an open boundary in the $x$ direction. 
This allows us to study the Fermi arc states in the $(k_y,k_z)$ plane. 
In the clean limit, the lattice Hamiltonian reads
\begin{widetext}
\begin{eqnarray}
	H^{\rm lat}(k_y,k_z)=\sum_{x}\left[
	h(k_y,k_z)
	|x\rangle\langle x| +
	U(k_y,k_z)
	|x+1\rangle\langle x| +
	U^{\dagger}(k_y,k_z)
	|x\rangle\langle x+1|\right]
	\label{eq:Hlat},
\end{eqnarray}
where 
\begin{eqnarray}
	h(k_y,k_z)=
	\begin{pmatrix}
	2-\cos(k_y)-2\cos(k_z)&-2\cos(k_y)&i\lambda e^{-i(\phi+\pi/4)}\sin(k_y)\\
	-2\cos(k_y)&-2+\cos(k_y)-2\cos(k_z)&-i\lambda e^{-i(\phi-\pi/4)}\sin(k_y)\\
	-i\lambda e^{i(\phi+\pi/4)}\sin(k_y) & i\lambda e^{i(\phi-\pi/4)}\sin(k_y)&0
	\end{pmatrix},
\end{eqnarray}
\begin{eqnarray}
U(k_y,k_z)=
\begin{pmatrix}
	-1/2&1-\sin(k_y)&i\lambda e^{-i(\phi+\pi/4)}/2\\
1+\sin(k_y)&1/2&i\lambda e^{-i(\phi-\pi/4)}/2\\
-i\lambda e^{i(\phi+\pi/4)}/2 & -i\lambda e^{i(\phi-\pi/4)}/2&0
\end{pmatrix}.
\end{eqnarray}
\end{widetext}
$H^{\rm lat}(k_y,k_z)$ is a $3L\times 3L$ matrix while $L$ represents the system size in the x-direction.
Therefore, we can directly calculate the eigenspectrum and corresponding wavefunction using exact diagonalization techniques for the lattice Hamiltonian.

To investigate the effect of potential disorder, we add the following Hamiltonian to Eq.~\eqref{eq:Hlat}:
\begin{eqnarray}
	H_{\rm dis}^{\rm lat}(k_y,k_z)=\sum_{x}V(x,k_y,k_z)I_{3\times 3}|x\rangle\langle x|.
\end{eqnarray}
Here the disorder potential $V(x,k_y,k_z)$ is drawn independently for each lattice site in the x-direction from the uniform distribution $[-W/2,W/2]$. 
Note that $V(x,k_y,k_z)$ can be also chosen uniformly or independently in the $k_y$ and $k_z$ direction in the momentum space~\cite{Slager2017}.

To extract the Fermi arc states of top and bottom surfaces, we select the state  with the largest weight by
\begin{eqnarray}
\mathop{\arg\max}_i\sum_{o=1}^3
\sum_{x}
|\langle \psi_i(k_y,k_z)|x,o\rangle|^2,
\end{eqnarray}
where $\psi_i$ is the $i$th eigenvector at $k_y, k_z$, $o$ labels the pseudospin degree of freedom, $x=\left[L-d,L\right]$ for the bottom surface and $x=\left[1,d\right]$ for the top surface, where $d=3$.


\section{Surface Berry curvature and surface Berry curvature dipole}\label{app:bcd}

The Fermi arcs of TPFs carry surface Berry curvature $\Omega_{x}(\vec{k})$.
We have showed the surface BC for the Fermi arcs on the top surface of the slab in Fig.~\ref{fig:surfbc} for $E=0.3$ at $W=0, 1,2,4$. 
The Fermi arcs and BC for the bottom surface is also shown in Fig.~\ref{fig:surfbcb}. 

In our case the Fermi arcs also have nontrivial surface BC dipole, which can lead to the nonlinear Hall currents in second-order response to an external electric field.
In the presence of a driving in-plane electric field, $E_k=\Re \{\mathcal{E}_k \mathrm{e}^{\mathrm{i}\omega t}\},$ the dc and second harmonic generated~(SHG) currents in the second order nonlinear effect are described as $j^{(0)}_i = \sigma_{ijk} \mathcal{E}^*_j \mathcal{E}_k$ and $j^{(2\omega)}_i = \sigma_{ijk} \mathcal{E}_j \mathcal{E}_k,$ respectively.
The tensor indices $i, j, k$ span the 2D coordinates $y, z$ of a slab geometry, because we impose open boundary conditions along the x-direction of the TPF Hamiltonian. 
Following previous theoretical works~\cite{Low2015, Sodemann2015}, we know that when the frequency of the the driving electric field is much smaller than the resonant
frequency for optical transitions, the nonlinear Hall-like current density corresponding to $\sigma_{ijj} $ with $i \neq j$ is
\begin{eqnarray}\label{current}
\vec{j}^{(0)}&=&\frac{e^3 \tau}{2 \hbar^2 (1+\mathrm{i}\omega \tau)} \hat{x}\times \vec{\mathcal{E}}^* (\vec{ D} \cdot\vec{ \mathcal{E} } ),\\\label{current1}
\vec{j}^{(2\omega)}&=&\frac{e^3 \tau}{2 \hbar^2 (1+\mathrm{i} \omega \tau)} \hat{x}\times \vec{\mathcal{E}} (\vec{ D} \cdot\vec{ \mathcal{E} } ),
\end{eqnarray}
where $\vec{ D}$ is the 2D surface BC dipole
and $\tau$ is the relaxation time. 
\begin{figure}[tb]
	\includegraphics[width=0.4\textwidth]{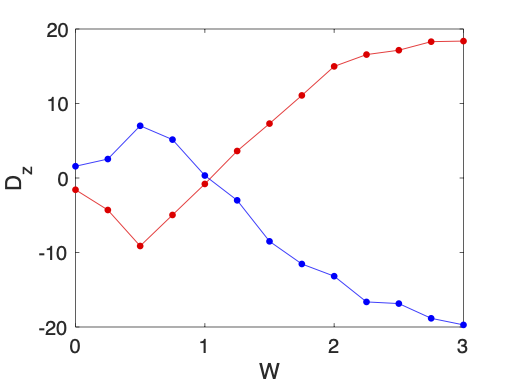}
	\caption{
	The surface Berry curvature dipole $D_z$ on the top/bottom surface as a function of disorder strength $W$. 
	Here we consider the Fermi energy at $E=0.3$, the thickness $ L=20$,
	and average $D_z$ over 400 disorder realizations.}
	\label{fig:bcd}
\end{figure}

We now discuss how the symmetries affect the BC dipole of the surface states considered in this work.
In two dimensions, the BC dipole is a pseudo-vector and it has been shown that the highest symmetry of a 2D crystal that allows for a finite $D_i$ is a single mirror line~\cite{Sodemann2015}. 
When the boundary is open along the x-direction of the model Hamiltonian, the slab has only one mirror symmetry in the z-direction, i.e. $H(-k_z)=H(k_z).$
Since this mirror symmetry enforces that $\Omega_{x}(k_y, k_z) = -\Omega_{x}(k_y, -k_z)$ in the surface Brillouin zone, it is evident that $D_{y}= 0$ while $D_{z}\neq0$. 
By using the stacked-layer construction, we numerically compute the surface BC dipole, setting the Fermi energy to $E=0.3$ and the thickness $L=20$, as shown in Fig.~\ref{fig:bcd}. 
Our result shows the the surface BC dipole has a non-monotonic dependence on disorder. 
 
{
	The surface BC dipole can be understood as follows. 
Eq.~(10) can be rewritten as $D_i=\int d^2 \vec{k}\delta(E-E_f)\frac{\partial E}{\partial{k_i}}\Omega_x(\vec{k})$ \cite{You2018}, where $\frac{\partial E}{\partial{k_i}}$ is the group velocity along the $i$ direction and $\Omega_x(\vec{k})$ is the BC.
 Turning on the disorder potential not only effectively varies the chemical potential and bands, but also induces hybridization between the surface states with the bulk states, which changes the localization of the wave function along the direction orthogonal to the surface and thus changes the value of the Berry curvature. 
 At the Fermi energy $E = 0.3$, the band dispersion of the surface state does not show a drastic change as a function of disorder. Thus, the major sign change in the BC dipole comes from Berry curvature, but not the group velocity. 
In Fig.~\ref{fig:surfbc}, it is evident that as disorder increases, the BC with opposite sign against that of the clean limit ($W=0$) grows substantially from two nodes. Thus, the BC dipole can change sign for a certain value of $W$. 
}
\bibliography{TI-references,references}

\clearpage

\end{document}